\documentclass[twoside,slac_one]{revtex4}
\usepackage{graphicx}
\usepackage{fancyhdr}
\usepackage{amsmath} % American Mathematics Society standards
\usepackage{bm}% bold math
\usepackage{amsxtra}
\usepackage{amssymb}
\usepackage{amsthm}
\usepackage{latexsym}
\usepackage{lscape}
\usepackage{subfig}
\usepackage[percent]{overpic}

\begin{document}
%\setpagewiselinenumbers
%\modulolinenumbers[5]
%\linenumbers

\title{A new part-per-million measurement of the positive muon lifetime and determination of the Fermi Constant}


\author{D. M. Webber, for the MuLan collaboration}
\affiliation{Department of Physics, University of Wisconsin-Madison, USA\footnote{This work was performed while the author was at the University of Illinois at Urbana-Champaign, USA}}

\begin{abstract}
The Fermi Constant, $G_F$, describes the strength of the weak force and is determined most precisely
 from the mean life of the positive muon, $\tau_\mu.$ Advances in theory have reduced the theoretical
 uncertainty on $G_F$ as calculated from $\tau_\mu$ to a few tenths of a part per million (ppm). Until
 recently, the remaining uncertainty on $G_F$ was entirely experimental and dominated by the uncertainty
 on $\tau_\mu.$ We report the MuLan collaboration's recent 1.0 ppm measurement of the positive muon lifetime.
 This measurement is over a factor of 15 more precise than any previous measurement, and is the most
 precise particle lifetime ever measured. The experiment used a time-structured low-energy muon beam
 and an array of plastic scintillators read-out by waveform digitizers and a fast data acquisition system 
to record over $2 \times 10^{12}$ muon decays. Two different in-vacuum muon-stopping targets were used in separate 
data-taking periods. The results from these two data-taking periods are in excellent agreement. The combined
 results give $\tau_{\mu^+}(\mathrm{MuLan})=2196980.3(2.2)$~ps. This measurement of the muon lifetime gives the most
 precise value for the Fermi Constant: 
$G_F(\mathrm{MuLan}) = 1.1663788 (7) \times 10^{-5}\ \mathrm{GeV}^{-2}$ (0.6~ppm). The lifetime is also
 used to extract the $\mu^-p$ singlet capture rate, which determines the proton's weak induced pseudoscalar
 coupling $g^{}_P$.
\end{abstract}

\maketitle

\thispagestyle{fancy}


\section{Introduction}

The Standard Model of particle physics has given an excellent description of particle interactions since it was first 
formulated in the early 1970's.  The predictive power of the Standard Model depends on its well-measured 
input parameters.  These parameters include the masses of the leptons and quarks and the mixing angles of the 
unitary CKM quark mixing matrix\footnote{Neutrinos are massless in the Standard Model.}.  The Standard Model also describes three of the fundamental forces of nature: the electromagnetic, weak, and strong forces.

The weak force requires three parameters to characterize its interactions.  The most precisely measured parameters are the Fermi Constant $G_F$, the mass of the Z-boson $M_Z$, and the fine structure constant $\alpha$.
The Fermi Constant describes the strength of the weak force, and is related to the electroweak gauge coupling g by
\begin{equation}
\frac{G_F}{\sqrt{2}} = \frac{g^2}{8M_W^2}\left( 1 + \Delta r \right),
\end{equation}
where $\Delta r$ includes 
%(the W propagator?) and 
all higher-order weak interaction loops.

The Fermi Constant is determined most precisely from muon decay.  Muon decay is most likely to proceed via the reaction 
$\mu^+ \rightarrow e^+ \nu_e \overline{\nu}_\mu.$
In Fermi theory, muon decay is described by a four-fermion contact interaction.  The relation between the muon lifetime and the Fermi Constant is
\begin{equation}
\frac{1}{\tau_\mu}= \frac{G_{F}^2 m_\mu^5}{192 \pi^3} \left(
1+\Delta q \right),  \label{eq:muondecay}
\end{equation}
where $\Delta q$ includes phase space, QED, hadronic, and radiative corrections.  In 1999, the 2-loop QED corrections were calculated, which reduced the theoretical uncertainty on the Fermi Constant as calculated from $\tau_\mu$ to less than 0.3~ppm; previously it was the dominant uncertainty \cite{vanRitbergen:All}.
This improvement in theory motivated  new measurements of the muon lifetime: MuLan \cite{MuLan:2007} and FAST \cite{FAST:2008}.

\section{Experiment Overview}

The Muon Lifetime Analysis (MuLan) experiment used a time-structured surface-muon beam and a symmetric detector to observe over $2 \times 10^{12}$ muon decays. 
 The experiment took place in the $\pi$E3 beamline at the Paul Scherrer Institute (PSI) in Villigen, Switzerland.  PSI has the most intense continuous proton beam in the world. 
 The $>2$~mA, 590 MeV proton beam
is directed through a 4-6~cm carbon target, producing pions.  The positive pions that decay at rest near the surface of the carbon target produce positive muons which are collected by the $\pi$E3 beamline and guided toward the MuLan experiment. 

A time structure of 5~$\mu$s beam-on and 22~$\mu$s beam-off 
is imposed on 
the continuous muon beam 
by a 25~kV electrostatic kicker \cite{kicker}.  During the 5~$\mu$s beam-on period, muons propagate down the beampipe and stop in a thin target at the center of the MuLan detector.  During the 22~$\mu$s beam-off measurement period, the muons are deflected into a beam collimator upstream of the experiment.
The beam extinction, defined by the ratio of beam-on to beam-off muon rate at the detector, is typically around 1000.

The muons stop in a target at the center of the MuLan detector.  
 The negative helicity of the muons is preserved during beam transport, and muons stop in the target with $\sim100\%$ polarization.  A slow precession of the ensemble muon spin from the Earth's magnetic field would introduce an oscillation in the observed muon decays which could pull the lifetime fit.  The muon stopping targets were chosen to dephase or precess the muon ensemble to remove or obviate the muon spin precession. 
During the 2006 run, the target was composed of a ferromagnetic alloy of 
$\approx60\%$ Iron, 
$\approx30\%$ Chromium,
and $\approx10\%$ Cobalt, called Arnokrome$^\mathrm{TM}$ III (AK-3) \cite{arnokrome}.  The 0.4-T internal magnetic field precessed the muon spins with period 18~ns.  This precession dephased the muon ensemble during the 5 $\mu$s beam-on period.  During the 2007 run, a crystalline quartz target with 130-Gauss externally-applied magnetic field was used.  In quartz, the majority of the muons form a muonium bound state consisting of a muon and an electron.  The magnetic moment of the paramagnetic bound state is $\approx200\times$ higher than that of the muon alone.  Muons in the paramagnetic state precess with period 2.6 ns, and muons in the diamagnetic state precess with period 550~ns.  The 550~ns muon precession period is observed.

\begin{figure}[ht]
\centering
\includegraphics[width=80mm]{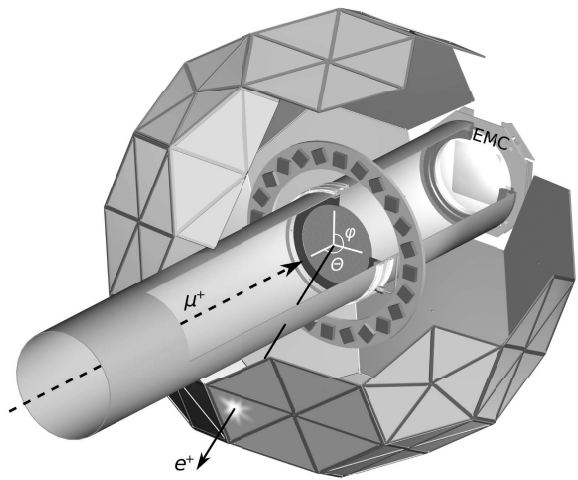}
\caption{The MuLan detector.  Muons are transported in vaccuum to a stopping target in the center of the detector.  The stopping target was composed of AK-3 in 2006 and crystalline quartz in 2007.  An external magnetic field was applied to the quartz target using a Halbach array of permanent magnets.  Both targets can open to allow the beam to pass to a wire chamber downstream of the experiment (EMC).} \label{fig:detector}
\end{figure}

The MuLan detector symmetrically surrounds the muon stopping target 
and records the outgoing positrons from muon decay (Fig. \ref{fig:detector}).
The detector is composed of 170 triangular scintillator tile pairs.  The coincident signal between inner and outer tiles of a pair defines a through-going particle.  The scintillation light from each tile is collected on one edge of each tile and guided toward a 29-mm photomultiplier tube (PMT).  The analog signal from each PMT is read-out by a 450 MHz waveform-digitizer (WFD).  When the signal goes over threshold (a ``hit''), 24 waveform samples are recorded along with the channel and the time.  Every hit is stored to tape, and $\approx60$~terabytes are required to store the information from $10^{12}$ muon decays.

\begin{figure}[ht]
\centering
\subfloat[ ]{
\includegraphics[width=80mm]{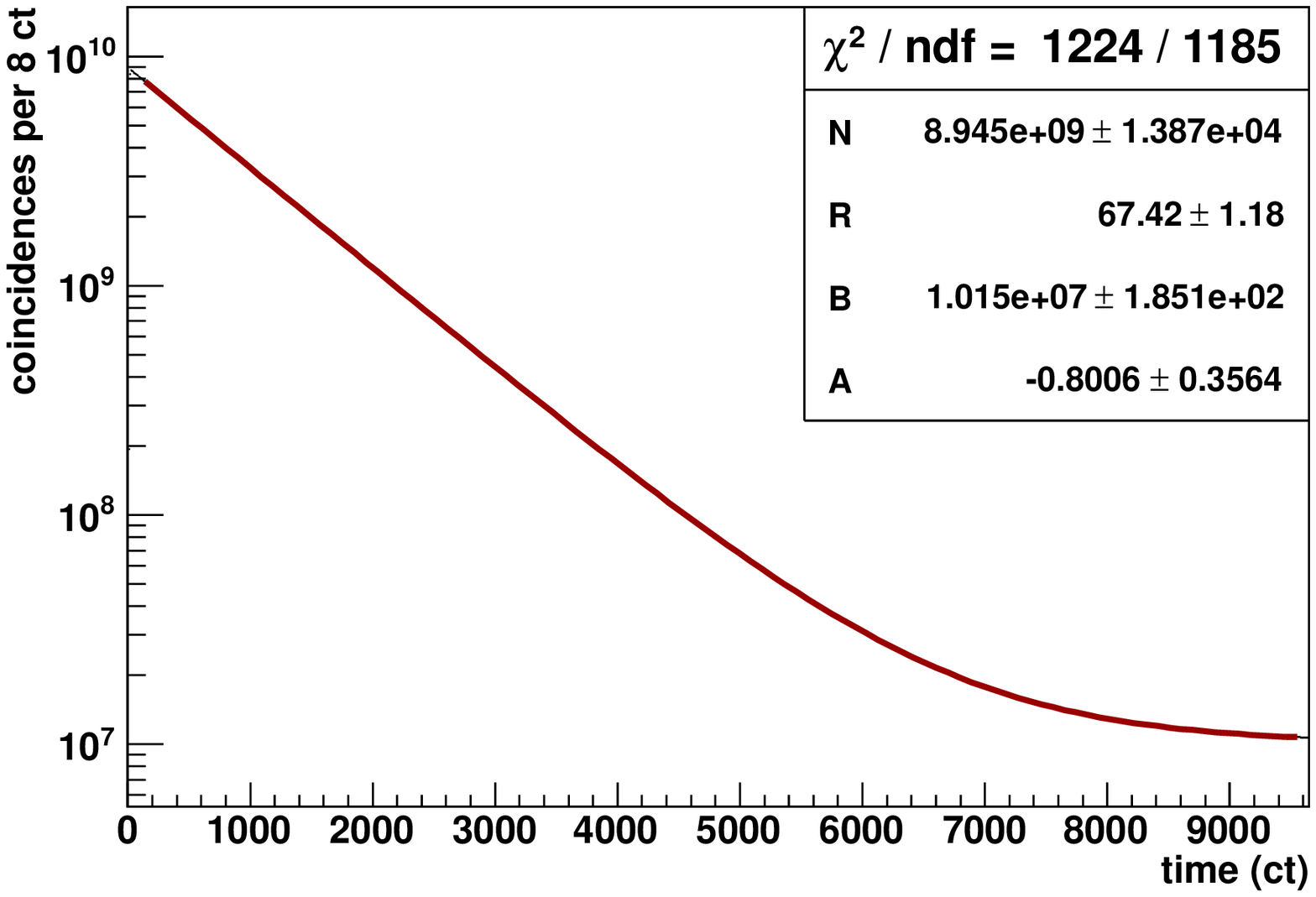}
\label{fig:fit}
}
\subfloat[ ]{
\includegraphics[width=80mm]{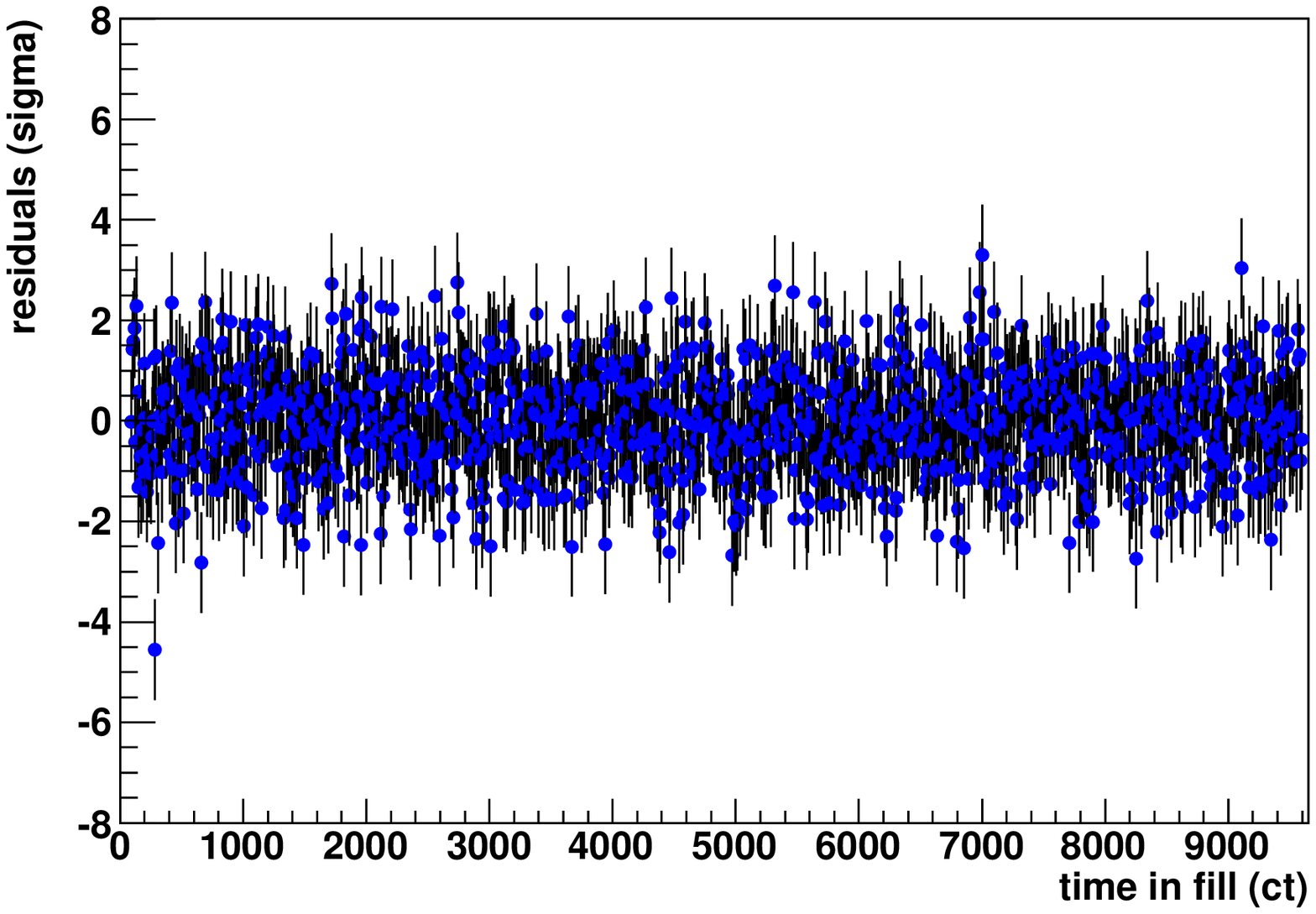}
\label{fig:residuals}
}
\caption{a) Coincident hits vs. time in measurement time for the 2006 dataset, fit with function \ref{eqn:lifetime}.  The reduced $\chi^2=1.03\pm0.04$ is acceptable.  b) The fit residuals show no structure.} 
\label{fig:fit+res}
\end{figure}

In the analysis, coincidences are identified as hits in the inner and outer tiles of a pair within a given time interval, usually 11~ns.  These coincidences are histogrammed versus time since the beginning of the measurement period (Fig. \ref{fig:fit+res}). This lifetime histogram is then fit with function 
\begin{equation}
  f(t) = (1 + A S(t)) N_0 e^{-t/\tau_\mu} + B,
  \label{eqn:lifetime}
\end{equation}
where $N_0$ is a normalization constant, $\tau_\mu$ is the muon lifetime, $B$ is a flat background, $S(t)$ is a small oscillation from the readout electronics (see section \ref{sec:gain}), and $A$ is the scale of the electronics oscillation.  During the analysis the lifetime was represented by $R$, defined as a ppm-difference from a reference value:
\begin{equation}
\tau_\mu = \tau_0 \left( 1 + R/10^6 \right).
\label{eqn:Rdef}
\end{equation} 

To prevent experimenter bias during the analysis, the WFD digitization frequency was blinded and the muon lifetime was known only in ``clock ticks'' (ct).  After the analysis of the systematic uncertainties was complete, the clock frequency was unblinded to reveal the measured muon lifetime.  

\section{Systematic Uncertainties}

During the 22-$\mu s$ measurement period (about 10 muon lifetimes), the hit rate in the detector changes by 3 orders of magnitude.  Any rate- or time-dependent change in detector performance during the measurement period could distort the fitted muon lifetime.  A list of systematic and statistical uncertainties is given in table \ref{tbl:ErrorTable}.  Two of these uncertainties, gain stability and pileup, will be discussed in detail.

\begin{table}[h!]
  \centering
  \caption{Systematic and statistical uncertainties in ppm units.
  The errors in different rows of the table are not correlated to
  each other.  Where only one error appears in a given row, the effect is $100\%$
  correlated between the two run periods.}
  \label{tbl:ErrorTable}
  \begin{tabular}{lcc}
    \hline
    Effect uncertainty in ppm       & ~~R06~~       &  ~~R07~~ \\
    \hline\hline
    Kicker stability                & ~~0.20~~  & ~~0.07~~ \\
    Spin precession / relaxation~~~~~    & ~~0.10~~  & ~~0.20~~ \\
    Pileup                          & \multicolumn{2}{c}{0.20} \\
    Gain stability                   & \multicolumn{2}{c}{0.25} \\
    Upstream muon stops             & \multicolumn{2}{c}{0.10} \\
    Timing stability                & \multicolumn{2}{c}{0.12} \\
    Clock calibration               & \multicolumn{2}{c}{0.03} \\
    \hline
    Total systematic                & 0.42 & 0.42 \\
    \hline
    Statistical uncertainty         & 1.14 & 1.68  \\
    \hline
    \hline
  \end{tabular}
\end{table}

\subsection{Gain Stability}
\label{sec:gain}

\begin{figure}[ht]
\centering
%\begin{overpic}[width=135mm,grid,tics=10]%
\begin{overpic}[width=135mm]%
{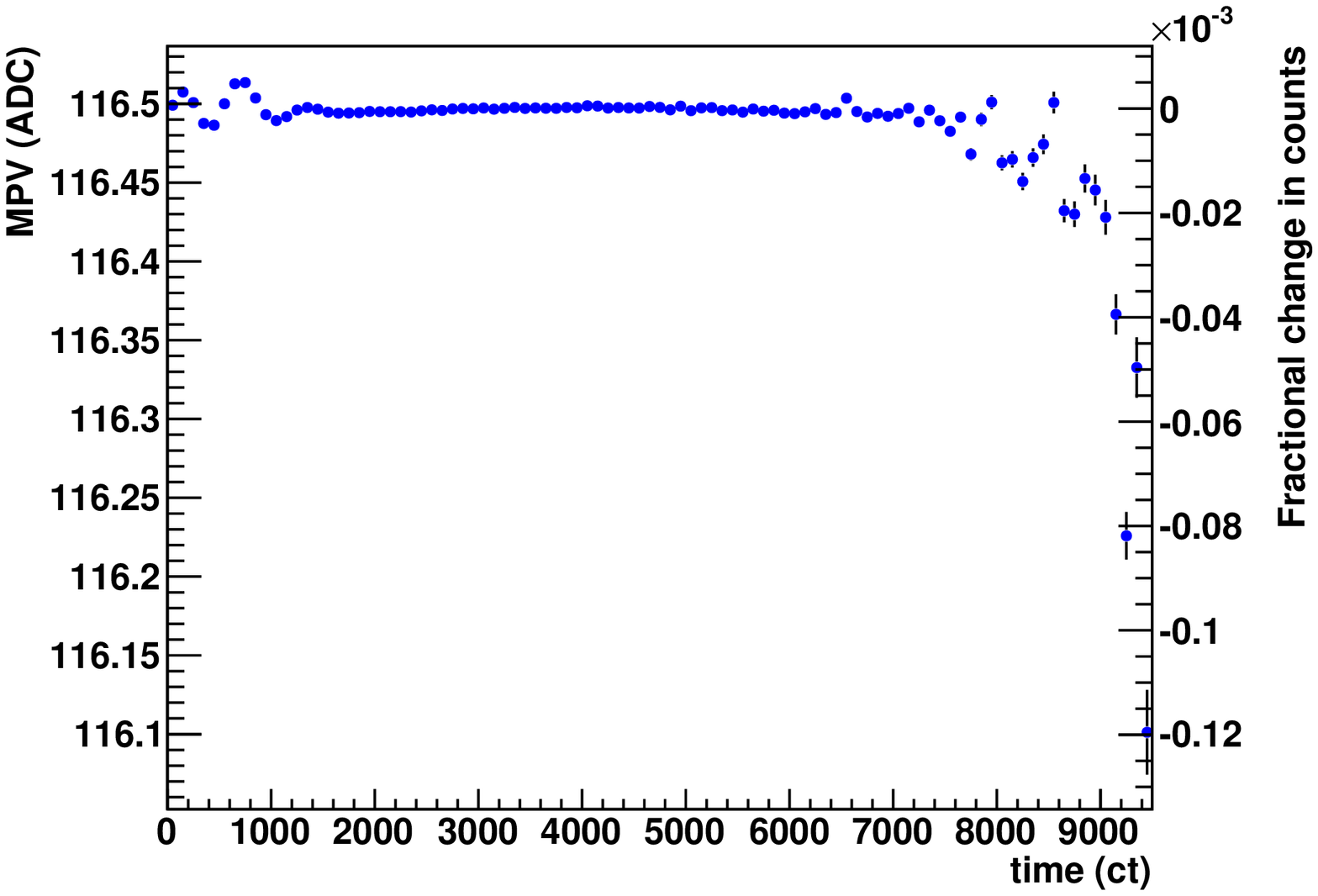}
  \put(20,10){\includegraphics[width=75mm]%
	  {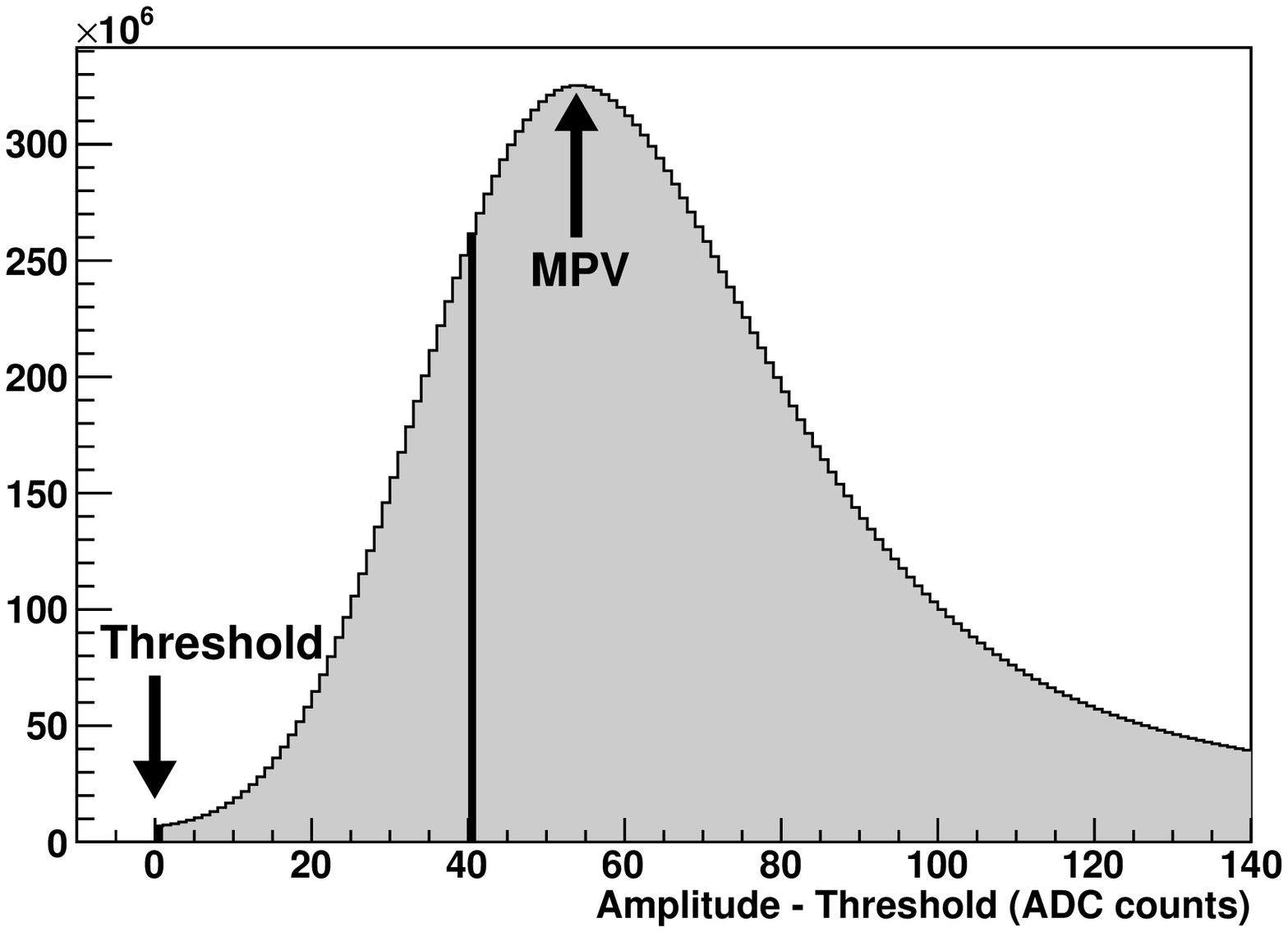}}
\end{overpic}
\caption{Most probable value for the pulse height (MPV) vs. time in measurement period.  The amplitude spectrum of the background, determined in the time range 9500-9600 ct (20.9-21.1 $\mu s$), was subtracted.  An electronics oscillation is visible early in the measurement period.  
Inset: amplitude minus threshold, summed over all detector PMTs.  At threshold, the fractional change in counts for a 1 ADC change in threshold is $\sim3.0\times10^{-4}$.  For comparison, at threshold+40 ADC, the fractional change in counts is $\sim114.4\times10^{-4}$.  
The effect on the lifetime histogram 
%change in count rate due to this oscillation wa
is found by scaling the oscillation in MPV by the fractional number of counts at the amplitude threshold.  Taking the oscillation into account shifts the fitted muon lifetime by $0.50\pm0.25$~ppm.
%For coincidences between tile pairs, the change in counts per change in threshold is doubled.
}
\label{fig:MPV}
\end{figure}

A change in signal height, or gain, will change the number of hits over threshold.  If the gain changes in time-dependent way it will introduce a perturbation on the lifetime histogram which may pull the fitted lifetime. For example, Fig. \ref{fig:MPV} shows a small oscillation in the pulse height most-probable-value early in the measurement period, which originates in the electronics.  Multiple analysis techniques were employed to characterize the effect of this oscillation on the fitted muon lifetime.  When the oscillation is taken into account, either by including it in the fit function or correcting the lifetime histogram, the fitted muon lifetime changes by $0.50\pm0.25$~ppm.

\subsection{Pileup}

Although the waveform digitizers (WFDs) provide complete waveforms for each hit, the analysis software cannot reliably distinguish two hits separated in time by less than 5 WFD samples ($\approx11$~ns).  When two hits occur within this so-called deadtime (DT), the second hit is lost.  This effect is called pileup. 

\begin{figure}[ht]
\centering
\includegraphics[width=135mm]{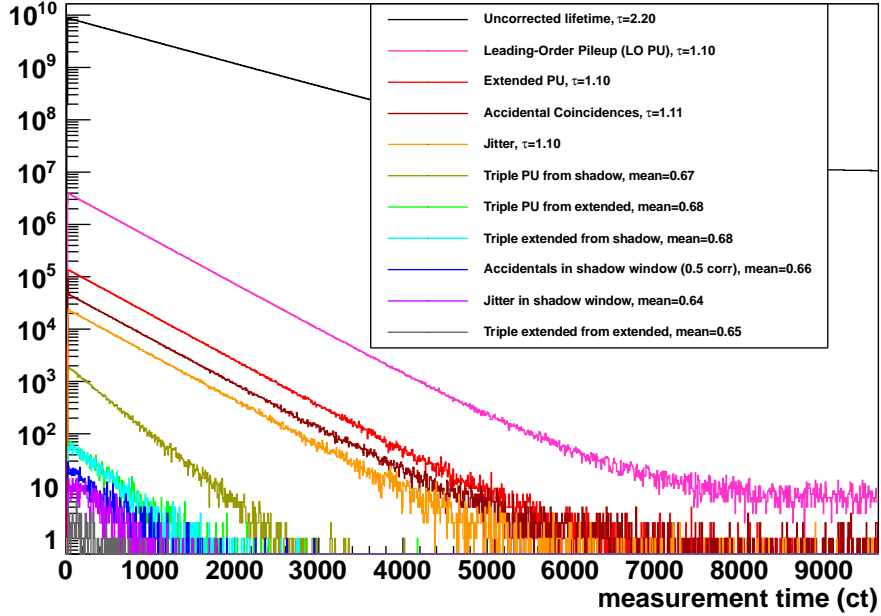}
\caption{Pileup correction histograms. Even negligibly small $\mathcal{O}(10^{-10}$) effects were considered.  Note that lifetime fits are approximate.}
\label{fig:AllPileup}
\end{figure}

Pileup is statistically corrected using the data.   When a hit is observed at time $t_1$ in beam fill $i$, a search is triggered in the next
 beam fill $i+1$ from time $t_1$ to $t_1+\mathrm{DT}$.  If a hit occurs in the same channel in this search window, it is stored in 
a pileup-correction histogram.  The pileup-correction histogram is added to the main lifetime histogram prior to fitting.
% for the muon lifetime. 
 This ``shadow-window'' method statistically corrects for the pileup by counting certain hits twice to compensate for the hits which are lost. 
 Several higher-order pileup effects were also considered, including triple-hits, accidental coincidences, and ``softening'' of the coincidence window boundaries due to jitter in the inner-outer coincidence times (Fig. \ref{fig:AllPileup}).

\begin{figure}[ht]
\centering
\subfloat[ ]{
\includegraphics[width=80mm]{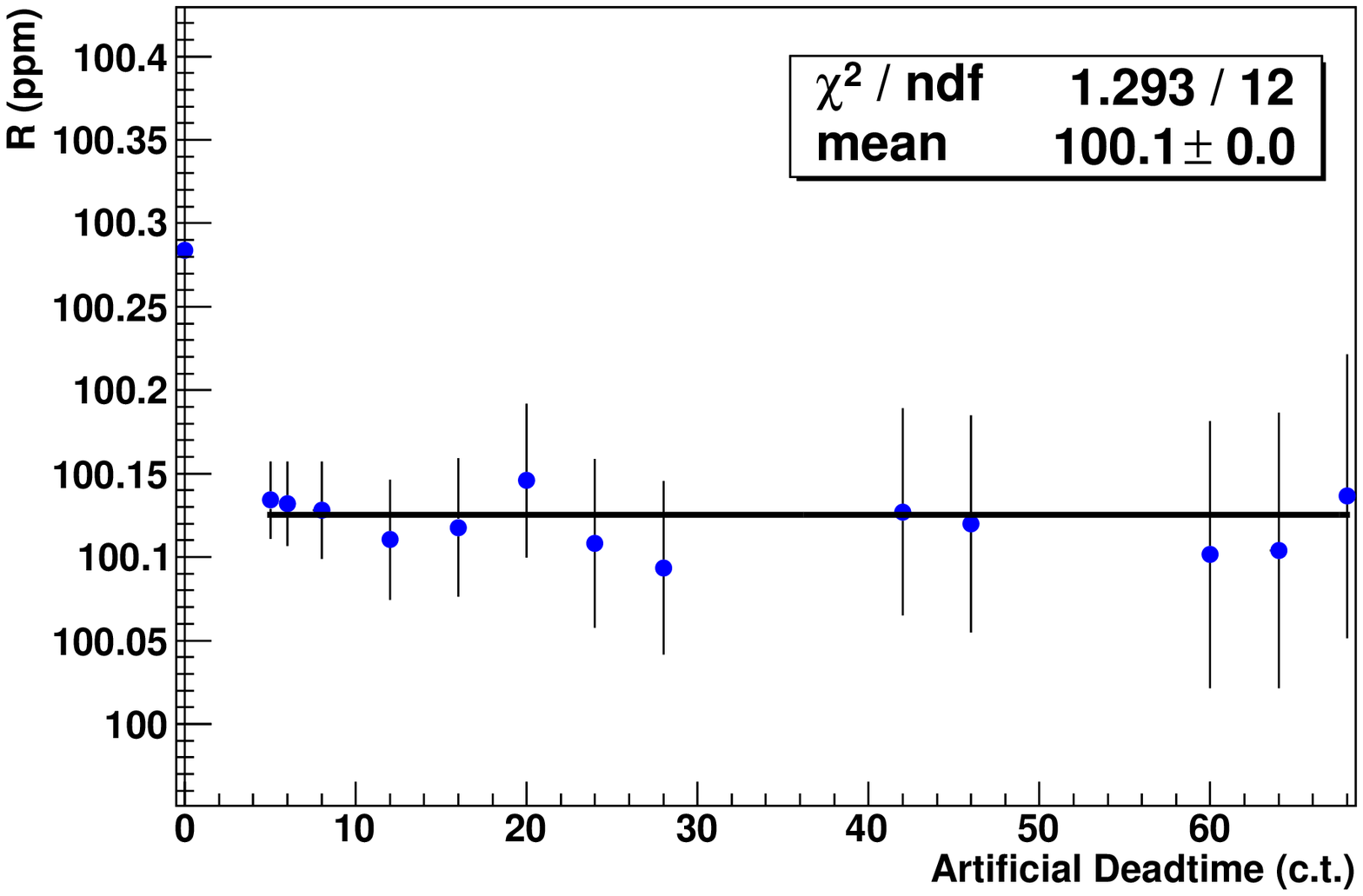}
\label{fig:R_vs_ADT_MC}
}
\subfloat[ ]{
\includegraphics[width=80mm]{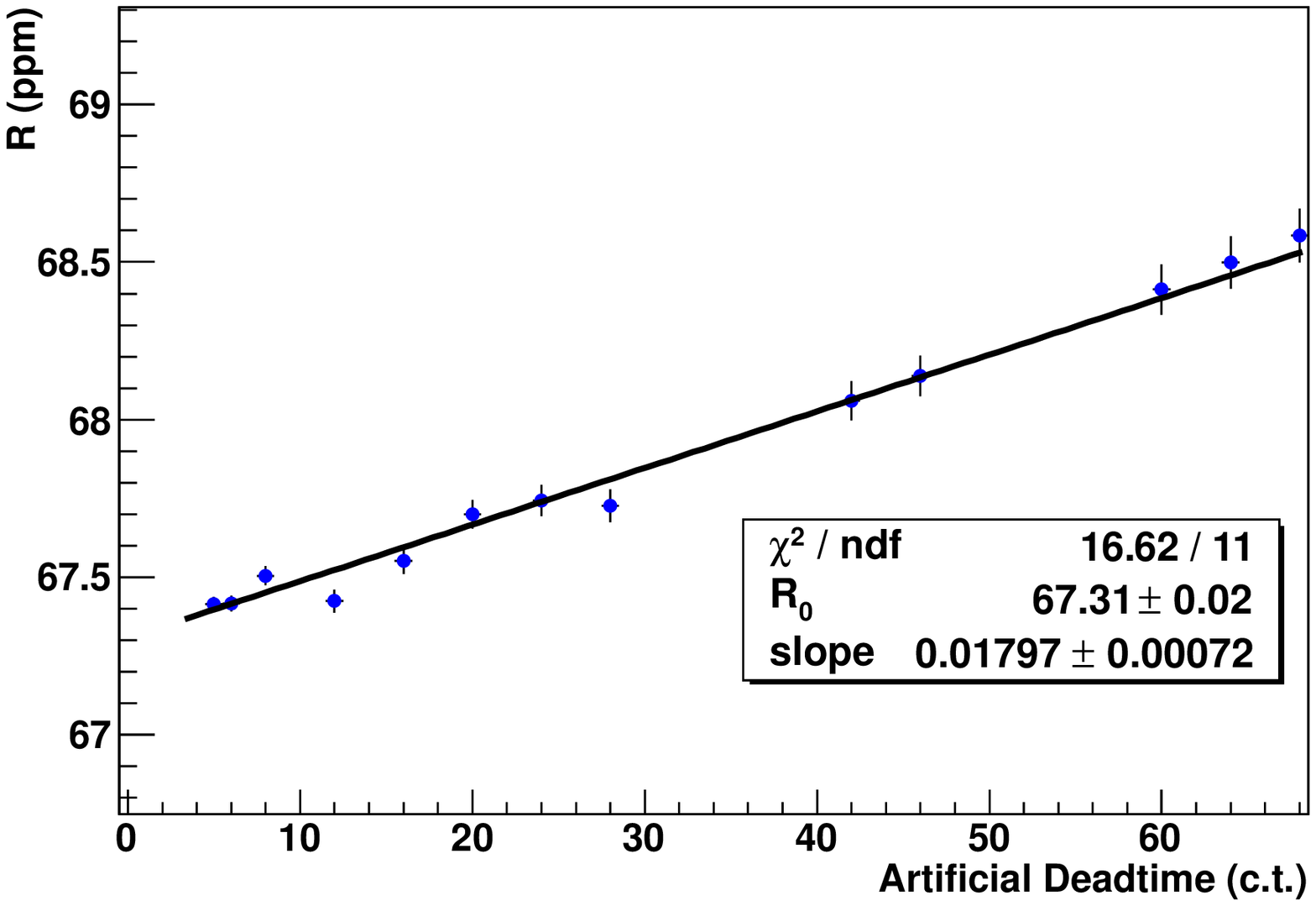}
\label{fig:R_vs_ADT_data}
}
\caption{a) Monte-Carlo pileup reconstruction for $10^{12}$ simulated muon decays. The simulated
lifetime R value is shown at the left for ADT=0 with statistical error bars 1.19 ppm,
larger than the range of the plot. The other points at selected ADTs are shown with error
bars allowed by the deadtime correction. The difference between the simulated lifetime and
the average reconstructed value is 0.16 ppm.
b) R vs. ADT for the 2006 dataset, after all deadtime corrections have been applied. The error bars
are the allowed deviations in R from the statistical deadtime reconstruction, and are smaller
than the overall statistical uncertainty of 1.14 ppm.
}
\end{figure}

The full pileup correction method was tested in Monte-Carlo with $10^{12}$ simulated muon decays (Fig. \ref{fig:R_vs_ADT_MC}).  The correction shows excellent consistency with the simulated lifetime over a range of artificially applied deadtimes (ADTs).  When this technique is applied to the data (Fig. \ref{fig:R_vs_ADT_data}), a residual slope exists of order 0.008 ppm/ns of applied deadtime, corresponding to a $\sim0.1\%$ pileup undercorrection.  This undercorrection is attributed to fluctuations in the PSI proton beam, resulting in small variations in muon rate from one beam fill to the next.  The extrapolation to ADT=0 gives the true muon lifetime. 

\section{Results}

\begin{figure}[ht]
\centering
\includegraphics[width=80mm]{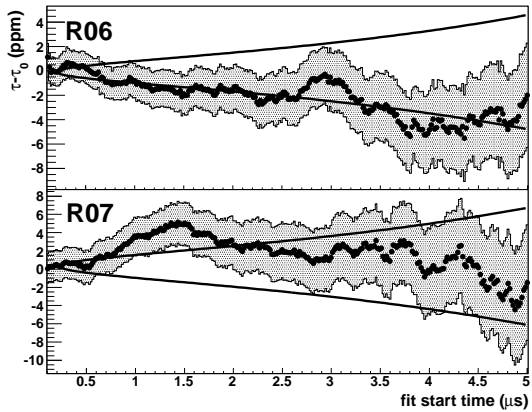}
\caption{Consistency of lifetime fit vs. fit start time.  Statistical errors are shown for each point, and the bands show the 1-sigma allowed variation.  No systematic trends are seen in either the 2006 or 2007 run periods.
}
% when varying the fit start time by over 2 muon lifetimes.} 
\label{fig:startR}
\end{figure}

In addition to the lifetime fit described by equation \ref{eqn:lifetime} and shown in Fig. \ref{fig:fit+res}, an exhaustive set of cross-checks was performed.  One of the most powerful cross-checks is a plot of the fitted muon lifetime vs. the fit start time.  As the start time of the fit is increased, any potential early-time effects are excluded from the fit.  A trend in the plot of lifetime vs. fit start time is an indication of an undiscovered systematic effect.  This plot is shown in Fig. \ref{fig:startR} for the 2006 and 2007 run periods.  The error bars are statistical, and the bands show the allowed 1-sigma statistical variation in the fit.  No systematic trends are seen.

\begin{figure}[ht]
\centering
\includegraphics[width=80mm]{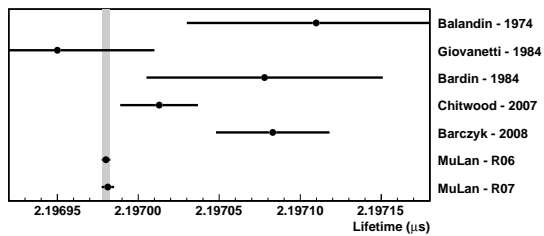}
\caption{Precision muon lifetime measurements.  Results from the 2006 and 2007 MuLan run periods are plotted separately to show their consistency.  The gray band shows the combined MuLan result \cite{lifetimes}.}
 \label{fig:plot_lifetime_history}
\end{figure}

During data analysis, the muon lifetime was known only in clock-tick units, and the blinded clock frequency was different for the two running periods.  After all systematic cross-checks were complete, the electronics clock frequency was unblinded to reveal the measured muon lifetimes:
\begin{eqnarray}\nonumber
\tau_{\mu}({\rm R06})&=& 2196979.9 \pm 2.5 \pm 0.9 {\rm ~ps}, \\
\tau_{\mu}({\rm R07})&=& 2196981.2 \pm 3.7 \pm 0.9 {\rm ~ps}.
\end{eqnarray}
where the first uncertainties are statistical and the second are systematic.  The results from the two running periods are in excellent agreement.  Combining these results yields
\begin{equation}
\tau_{\mu}({\rm MuLan}) = 2196980.3 \pm 2.2 {\rm ~ps~~~(1.0~ppm)}.
\label{finalresult}
\end{equation}
These results are shown in the context of previous precision muon lifetime experiments in Fig. \ref{fig:plot_lifetime_history}.

From this value for $\tau_\mu$, we derive a new value for $G_F$ following equation \ref{eq:muondecay}:
\begin{equation}
G_F({\rm MuLan}) = 1.1663788(7) \times 10^{-5} {\rm
~GeV^{-2}}~~~({\rm 0.6~ppm}),
\end{equation}
where the dominant contribution to the $G_F$ uncertainty is 0.5~ppm from the muon lifetime.

\begin{figure}[ht]
\centering
\subfloat[ Using previous $\tau_{\mu^+}$ world average ]{
%\subfloat[ ] {
\includegraphics[width=80mm]{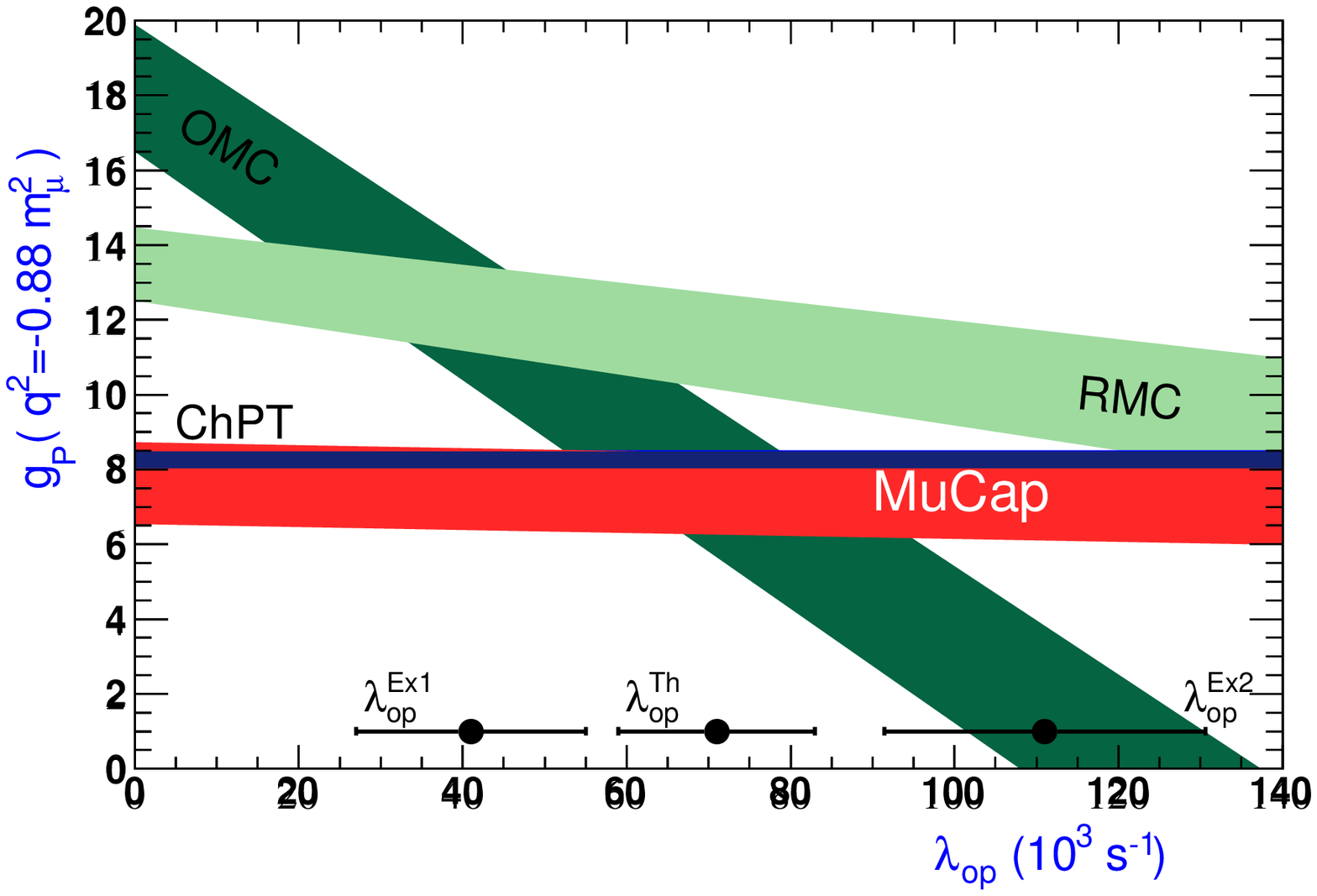}
\label{fig:sc_old}
}
%\subfloat[ ]{
\subfloat[ Using MuLan result for $\tau_{\mu^+}$ ]{
\includegraphics[width=80mm]{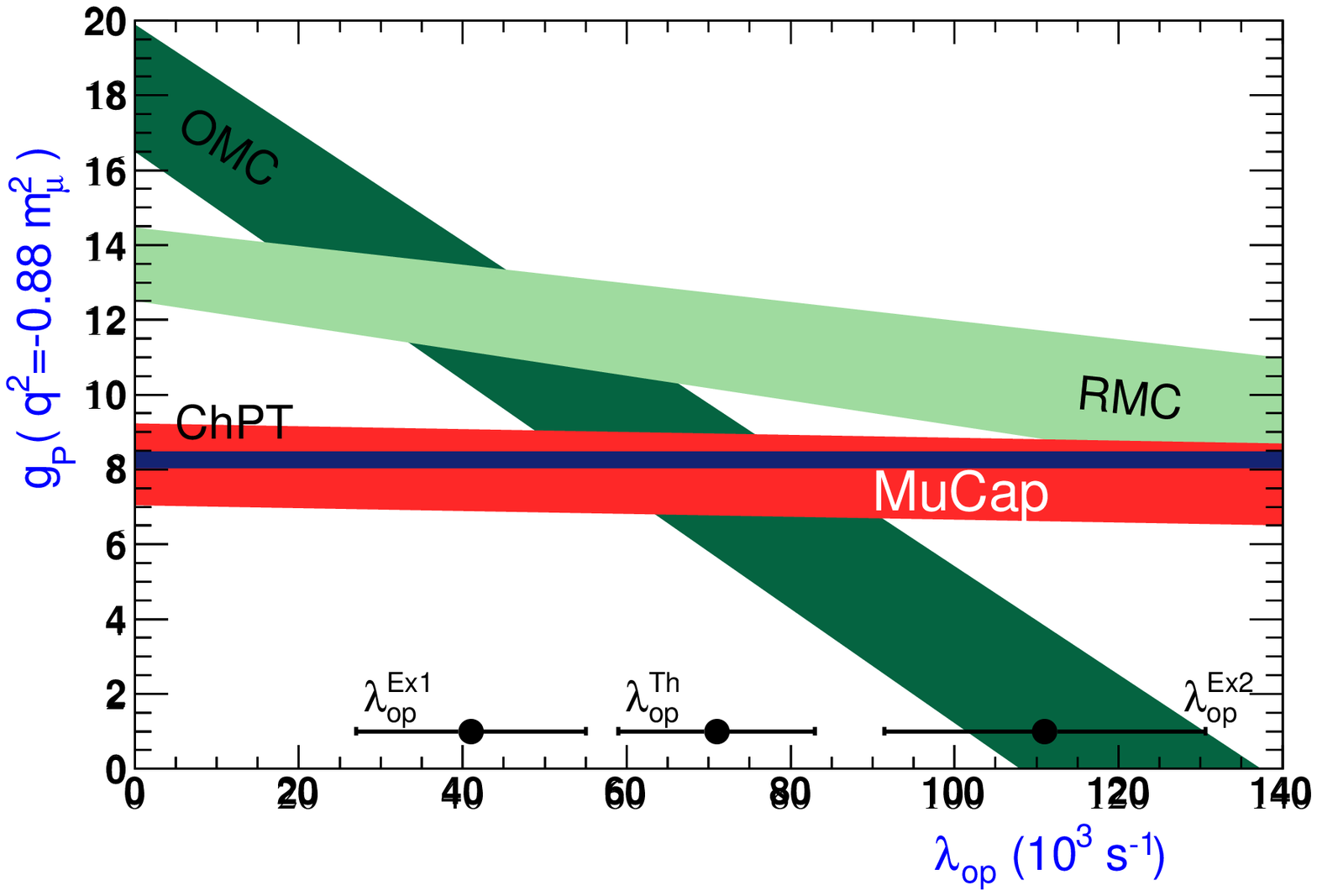}
\label{fig:sc_new}
}
\caption{Results from the MuCap experiment, which measures the negative muon capture rate on the proton and derives the pseudoscalar form factor of the proton, $g^{}_P$.  A precise value of the positive muon lifetime is necessary to derive the capture rate from the measured negative muon lifetime in hydrogen.  Using the MuLan result shifts the derived value of $g^{}_P$ into better agreement with theory.  The parameter $\lambda_\mathrm{op}$ is the transition rate of muonic molecular hydrogen (p$\mu$p) from the ortho to the para state.  These ortho and para states have different muon capture rates.  MuCap is designed to be relatively insensitive to the transition rate, which allows a clear extraction of $g^{}_P$ from the measured muon capture rate.} 
\label{fig:sc}
\end{figure}

Aside from a new determination of $G_F$, $\tau_\mu^+$ is also used in muon capture experiments.  The MuCap experiment measures the lifetime of negative muons in protium gas \cite{MuCap:2007}.  The difference between the positive and negative muon lifetimes is used to extract the singlet capture rate for the process $\mu^-p\rightarrow\nu_\mu n$.  The capture rate is then used to derive the pseudoscalar form factor of the proton, $g^{}_P$.  Fig. \ref{fig:sc} shows that using the MuLan result for $\tau_\mu^+$ shifts the value for $g^{}_P$ into better agreement with theory.

\section{Summary}
The Muon Lifetime Analysis (MuLan) experiment has determined a new precise value for the positive muon lifetime, $\tau_\mu^+$.  The experiment measured over $10^{12}$ muon decays using a time-structured surface-muon beam and a symmetric detector.  The result, 
$\tau_{\mu}({\rm MuLan}) = 2196980.3 \pm 2.2 {\rm ~ps~~~(1.0~ppm)}$, is more than 15 times more precise than any previous measurement and is used to provide the most precise value for the Fermi Constant, 
$G_F({\rm MuLan}) = 1.1663788(7) \times 10^{-5} {\rm~GeV^{-2}}~~~({\rm 0.6~ppm})$
\cite{MuLan:2010}.
The muon lifetime is also used to extract the muon capture rate $\mu^-p\rightarrow\nu_\mu n$ and the pseudoscalar form factor of the proton, $g^{}_P$.

\begin{acknowledgments}
Support during this project was provided by the National Science Foundation
(NSF) under grant NSF PHY 06-01067, and the experiment hardware was largely funded by
a special Major Research Instrumentation grant, NSF 00-79735.
%, ``Collaborative Research:
%The MuLan Project - Development of instrumentation for a new high-precision determination
%of the Fermi coupling constant.''
The data analysis for this work was supported by the National Center for Supercomputing Applications
under research allocation TG-PHY080015N and development allocation PHY060034N
and utilized the ABE and MSS systems.

\end{acknowledgments}

\end{document}